\documentclass{ws-rv961x669}
\usepackage[square]{ws-rv-van}		
\usepackage{hyperref}				
\usepackage{cleveref}				

\makeindex							

\begin{document}

\chapter{Effective Field Theory of Quantum Black Holes}

\author[S. Choi and F. Larsen]{Sangmin Choi and Finn Larsen}

\address{Department of Physics and Leinweber Center for Theoretical Physics, \\University of Michigan, Ann Arbor, MI 48109-1120, USA}

\begin{abstract}
We review and extend recent progress on the quantum description of near-extremal black holes in the language of 
effective quantum field theory. With black holes in Einstein-Maxwell theory as the main example, we derive the Schwarzian low energy description of the AdS\textsubscript{2} region
from a spacetime point of view. We also give a concise formula for the symmetry breaking scale, we relate rotation to supersymmetry, and we discuss quantum corrections to black hole entropy.
\end{abstract}

\body

\section{Introduction}

This article was conceived as a written version of a presentation on original research addressed at physicists who do 
not necessarily follow current developments closely.\footnote{An extended version of a talk given at {\it Quantum Connections 2021}, a postpandemic workshop held June 21--25, 2021 in Stockholm, Sweden, honoring the life and science of Frank Wilczek.} To achieve this we have opted to focus on a single research direction, the thermodynamics of a black hole near its ground state. We review progress on the subject over the last few years in a manner that is self-contained and organized as an alternative to the prevailing narrative. In this context we address several gaps in the literature.

Central aspects of black hole physics are captured by focusing on two spacetime dimensions, essentially the radial evolution near the event horizon and its interplay with time. An important recent advance was the understanding that, at low energy, this near-horizon geometry shares its symmetry structure with certain quantum mechanical models in a single dimension \cite{Sachdev:1992fk,Sachdev:2010um,Kitaev,Rosenhaus:2018dtp}. Low energy effective field theory is entirely determined in terms of its symmetries so this construction offers an explicit model of holography, a radial direction ``emerges" in a model that is formulated entirely in terms of time. The one-dimensional boundary theory obscures the original black hole interpretation, of course, but it can be recovered by comparing its low energy observables with a gravitational theory in two dimensions that has the same symmetry breaking pattern. For example, Jackiw-Teitelboim gravity (and its relatives) has proven instructive and so it has become a canonical benchmark \cite{Maldacena:2016upp}. 

The approach we present in this paper is improved, because it is constructive, and is at any rate distinct and complementary. We analyze the symmetry of the near-horizon region {\it ab initio}, without reference to a specific bulk Lagrangian. In this set-up we are able to explicitly identify the physical degrees of freedom directly in spacetime and, because there are no mysterious divergences, there is no dependence on unknown UV physics. 
Our strategy is possible because, rather than focusing on the radial mode, we identify the boundary theory as the effective Lagrangian of the complex structure in the bulk geometry. This maneuver, technical as it may appear, is satisfying because it is the natural adaption to two dimensions of the standard AdS/CFT correspondence in higher dimensions, except that in two dimensions we can be much more explicit. 

With the Lagrangian in hand, we determine the scale of the low energy theory by comparing with the geometry slightly beyond the near-horizon region. This computation follows the standard UV matching procedure in effective field theory precisely and adapts the nAttractor mechanism established in \cite{Larsen:2018iou} to the black holes considered as the main examples here. We also show quantitatively that ${\cal N}=2$ supersymmetry, previously developed somewhat formally \cite{Fu:2016vas,Forste:2017apw}, can be interpreted physically as a relation between rotation of the black hole and thermal excitations away from the supersymmetric limit. 

The recent progress we discuss, including the refinements we introduce, rely entirely on low energy effective field theory. Importantly, the underlying principles
of nonlinearly realized symmetry, realized by an effective Lagrangian, apply also in the quantum theory. Therefore, the quantum corrections to the low-temperature partition 
function, determined as $\sim T^{3/2}$ by many researchers, must agree with appropriate complementary descriptions that apply at the highest and lowest energies
where the ``low temperature" description is justified. We argue that the limitation at high energy is due to Kaluza-Klein modes on the sphere, {\it ie$.$} higher partial waves of spacetime fields,
and at very low energy it is due to finite volume effects. These complementary regimes are themselves described by effective quantum field theories, albeit different ones, that do not depend 
on the UV completion of quantum gravity. Therefore, at energies where regimes of applicability overlap, quantum corrections must agree. In this presentation, we review results on the quantum corrections to black hole entropy, at zero and generic temperatures, and adapt them to establish the expected agreements. This serves to illustrate the power of effective quantum field theory, as well as its limitations.

This paper is organized as follows. In section \ref{sec:BHs} we introduce the basic thermodynamics of Kerr-Newman black holes in $D=4$ spacetime dimensions, with focus on their near-extremal behavior. In section \ref{nhsymm} we discuss the near-horizon geometry of the black hole. Specifically, we determine the space of distinct geometries by analyzing the applicable large diffeomorphisms. In section \ref{sec:effthy} we leverage this data to construct the effective action of low energy excitations of the geometry. 
In section \ref{sec:ssb} we discuss the scales of the spontaneously broken symmetry which, we show, enjoy some protection due to supersymmetry. Moreover, the two scales in the problem are related to one another by supersymmetry. In section \ref{sec:quantum} we discuss the power and limitations of effective quantum field theory for gravity, by comparing the effective description of the near-horizon geometry with computations of the quantum corrections to black hole entropy in various regimes. We finish, in section \ref{sec:summ}, with a brief summary aimed at further applications of the spacetime wave functions we determine.

\section{Black Holes in Einstein-Maxwell Theory}
\label{sec:BHs}
In order to keep the presentation broadly accessible, as our primary setting 
we consider $D=4$ Einstein-Maxwell theory. The black holes in this theory are the Kerr-Newman black holes. They are asymptotically flat and depend
on just three quantum numbers: mass $M$, angular momentum $J$, and electric charge $Q$.\footnote{Black holes in $D=4$ Einstein-Maxwell theory can also have magnetic charge $P$. Because of classical duality symmetry, their geometries depend only on the combination $\sqrt{Q^2 + P^2}$, and so taking $P=0$ involves no loss of generality at the classical level. In the quantum theory there are interesting subtleties (including \cite{Coleman:1991ku}) but they are not closely connected to the main points of this presentation. }

The Bekenstein-Hawking area law, applied to the Kerr-Newman black holes, gives the entropy:\footnote{In this equation we display $G_4$, in order to record dimensions once and for all, but henceforth we generally take $G_4=1$ and restore it only when needed for clarity.} 
\begin{equation}
S = \frac{A}{4G_4} = \frac{\pi}{G_4} \left[ \frac{J^2}{M^2}  +  \left( MG_4 + \sqrt{ (MG_4)^2 - Q^2 - \frac{J^2}{M^2}} \right)^2 \right] ~.
\end{equation}
The argument of the square root is negative if, for given charges $J$, $Q$, the mass violates the extremality bound 
\begin{equation}
M^2 \geq M^2_{\rm ext} = \frac{1}{2}Q^2 + \sqrt{\frac{1}{4} Q^4 + J^2} ~.
\label{eqn:extbound}
\end{equation}
The pathology of the entropy formula when the black hole mass is too small is a genuine feature of the underlying black hole solutions: regular geometries exist if and only if 
$M\geq M_{\rm ext}$~.

The basic $D=4$ Einstein-Maxwell theory we focus on, a theory that is well-known from elementary textbooks, is identical to the 
bosonic part of ${\cal N}=2$ minimal supergravity. In the supersymmetric theory there are also two gravitini but fermions can only appear in the action at quadratic (or higher) order so, 
as far as bosonic solutions to the theory are concerned, it is consistent to set them to zero. From this point of view the significance of supersymmetry is that there are {\it two} independent supercharges in ${\cal N}=2$ supergravity and, when they act sequentially, they relate gravity and the electromagnetic field. 

The inequivalent orderings of the  two supersymmetries is controlled by the supersymmetry algebra. In
a unitarity representation, it yields the BPS bound: 
\begin{equation}
\{ {\cal Q}, {\cal Q}^\dagger \} = 2 ( M  ~ - ~ Q ) ~\geq ~0  ~.
\label{eqn:SUSYbound}
\end{equation}
The notation here is schematic: we suppress the spinor indices on the supersymmetry generators ${\cal Q}$ and also suppress their two-fold multiplicity, but the
numerical coefficients are accurate. Comparison with the extremality condition \eqref{eqn:extbound} shows that the supersymmetry condition is stronger, its saturation 
corresponds to {\it two} conditions:
\begin{equation}
M = Q~~~~ {\bf and}~~~ J=0~.
\label{eqn:twocond}
\end{equation}
The broader lesson, applicable in much more general settings, is that supersymmetric black holes are extremal (for the given charges) {\it and} their charges
must satisfy a constraint. In Einstein-Maxwell theory the constraint on charges is very simple, just $J=0$, but for supersymmetric black holes with asymptotically AdS-geometry it can be much more complicated \cite{Kim:2006he,Hosseini:2017mds,Choi:2018hmj}.

The supersymmetric black holes \eqref{eqn:twocond} form a one parameter family among the general Kerr-Newman black holes with quantum numbers $(M, J, Q)$. 
{\it Nearly} supersymmetric black holes depend on the reference BPS black hole and, in addition, on two continuous real parameters that break supersymmetry. 
Concretely, the mass of a nearly supersymmetric black hole exceeds the supersymmetric bound due to temperature and/or angular velocity: 
\begin{equation}
\label{eqn:massform}
M ~~ =  ~~ Q ~~ + ~~ \frac{1}{2} Q^3 \Big( (2\pi T)^2+\Omega^2\Big)~.
\end{equation}
The temperature must be nonnegative while the angular velocity can have either sign. 

The scale of symmetry breaking is \cite{Preskill:1991tb}
\begin{equation}
\frac{C_T}{T} = 4\pi^2 Q^3~.
\label{eqn:SSBscale}
\end{equation}
The specific heat $C_T$ is proportional to temperature $T$ in the nearly supersymmetric regime so the ratio $C_T/T$ is useful as a symmetry breaking scale that characterizes the state of black hole matter \cite{Preskill:1991tb}. For a macroscopic black hole it is much larger than the horizon size $Q^3  G_4 \gg Q$. This hierarchy of scales expresses the collective nature of low-lying black hole excitations by introducing an energy scale that is much lower than for a particle in a box with comparable size. In string theory constructions this ``large string scale" is realized by the level of an ${\cal N}=2$ superconformal algebra in two dimensions and is closely related to spectral flow \cite{Maldacena:1996ds,Cvetic:1998xh,Dijkgraaf:2000fq,Kraus:2006nb}. 

Summarizing so far, there are several important symmetries in Einstein-Maxwell theory,  and they have an interesting interplay: 
\begin{itemlist}
\item
The supersymmetric ground state is very special and imposes two conditions \eqref{eqn:twocond}. It realizes superconformal symmetry. 

\item
There are two distinct ways to break superconformal invariance: finite temperature and/or rotational velocity (finite angular momentum). 

\item
The two breaking mechanisms are characterized by a single symmetry breaking scale \eqref{eqn:SSBscale} because the two potentials are related by supersymmetry. 
\end{itemlist}

\section{Near-Horizon Symmetry}
\label{nhsymm}

In general relativity gravity is a theory of coordinate transformations (diffeomorphisms) so,
to understand all the symmetries of the physical system, we must study the black hole geometry.  
The near-horizon geometry of an extremal 4D black hole is AdS$_2\times S^2$ \cite{Kunduri:2007vf}. 
At very low energy, excitations on the sphere are suppressed but nontrivial dynamics is encoded in the AdS$_2$ factor. 
We introduce complex coordinates and present the metric as
\begin{equation}
ds^2_2  =  g_{\mu\nu} dx^\mu dx^\nu = \frac{4}{(1-|z|^2)^2} dz d{\bar z}~. 
\label{eqn:zzbarPoincare}
\end{equation}
Here $z=0$ is the black hole horizon (in Euclidean signature), the radial evolution 
along $0\leq |z|<1$ represents the throat at finite spatial distance from the horizon, and the apparent singularity at $|z|=1$ is infinitely far from the horizon, where
the black hole ``attaches" to the surrounding space. 

The geometry \eqref{eqn:zzbarPoincare} is a representative of the black hole ground state but it is not unique, there is an entire manifold of degenerate vacua that can be
constructed by acting with symmetries. As candidates for alternative vacua, consider the metric deformations: 
\begin{equation}
\delta g_{zz} = ~~z^{n-2}~~, ~n=2, 3, \cdots~.
\label{eqn:dekgzz}
\end{equation}
These variations are all {\it normalizable}: their square as complex functions, computed by contracting indices using the metric read off from \eqref{eqn:zzbarPoincare}, 
give a finite result when integrated over the entire conformal disk $|z|<1$. This happens because the diverging volume factor $\sqrt{g}$ at the boundary $|z|\to 1$ is dominated by index contractions with the factor $g^{z{\bar z}}$ that drops off rapidly as $|z|\to1$. 

Before concluding that the metric deformations \eqref{eqn:dekgzz} represent low-lying excitations of the black hole, recall that metric deformations generated by coordinate transformations are not genuinely different geometries, the configuration space of gravity is the space of geometries up to coordinate transformations. In other words, diffeomorphisms $\delta x^\mu = \xi^\mu$ transform the metric coefficients as
\begin{equation}
\delta g_{zz} = 2\nabla_z \xi_z   ~,
\label{eqn:gzzdiff}
\end{equation} 
but the corresponding geometry is physically identical, it has simply been recast in a new coordinate system. 

The redundancy realized by diffeomorphisms is huge, especially in low spacetime dimensions. In particular, the sample deformation of the metric \eqref{eqn:dekgzz} is generated from the original ``reference" geometry \eqref{eqn:zzbarPoincare} 
by the coordinate transformation \eqref{eqn:gzzdiff} with
\begin{equation}
\label{eqn:xin}
\xi^{\bar z}  = \frac{1}{4} \left( \frac{1}{n-1} z^{n-1}  - \frac{2{\bar z}}{n} z^n   + \frac{{\bar z}^2}{n+1} z^{n+1}\right)  - \frac{1}{2(n-1)n(n+1)} {\bar z}^{n+1}~. 
\end{equation} 
%
%
%

The last term in this equation is such that the diagonal components $g_{z{\bar z}}$ are invariant. 
If taken at face value, this shows that the deformations $\delta g_{zz} = ~~z^{n-2}$ are pure diffeomorphisms, {\it ie$.$} they change the metric but not the geometry. 
However, the diffeomorphism $|\xi^{\bar z}| \sim 1$ near the boundary $|z|  = 1$ so it is {\it not} normalizable: the square of the vector $\xi^{\bar z}$ must be computed 
using the background conformal factor $g_{z{\bar z}}$ which diverges, and then integrated over the volume with divergent measure $\sqrt{g}$. Therefore, the variations \eqref{eqn:dekgzz} are {\it not} pure diffeomorphisms. In other words, since they are normalizable, and not generated by coordinate transformations, they {\it are} physical metric deformations \cite{Camporesi:1994ga,Banerjee:2010qc,Sen:2012kpz,Larsen:2014bqa}.

The modes \eqref{eqn:gzzdiff} are a basis for all configurations that can appear in this way. 
We can combine them (all integers $n=2,3,\ldots$) into a single function $\epsilon(z)$ that is regular at the origin $z=0$
where $\epsilon = \partial_z \epsilon =\partial^2_z \epsilon=0$ and parametrizes the generators of diffeomorphisms through 
\begin{eqnarray}
\xi^{\bar z} & = & {\bar \epsilon}({\bar z})- \frac{1}{2} \int_0^z (1 - z'{\bar z})^2 \partial^3_{z'} \epsilon(z') dz'
\nonumber\\
& = & {\bar \epsilon}({\bar z})- \frac{1}{2}(1 - z{\bar z})^2 \partial^2_{z} \epsilon(z) - {\bar z}(1 - z{\bar z}) \partial_{z} \epsilon(z)  - {\bar z}^2\epsilon(z) 
~, 
\label{eqn:inftrans}
\end{eqnarray}
and its complex conjugate. These transformations all preserve conformal gauge, the diagonal metric displayed in \eqref{eqn:zzbarPoincare} is invariant, and 
so they act on the geometry only by adding simple off-diagonal components: 
\begin{equation}
\label{eqn:hzzinf}
\delta ds^2 = -2 \left( \partial^3_z \epsilon(z) dz^2  + {\rm c.c.} \right)~.
\end{equation}
Thus the space of distinct geometries is parametrized by 
\begin{equation}
ds^2 = h_{zz}(z)dz^2 + {\rm c.c.} ~, 
\label{eqn:qhd}
\end{equation}
where the function $h_{zz}(z)$ is regular at the origin. The regular metric deformations that are not generated by diffeomorphisms are called quadratic holomorphic differentials (QHD's).

The analysis of AdS$_2$ presented here is reminiscent of string theory in the world-sheet formalism but there are important differences \cite{Charles:2019tiu}. The reference metric \eqref{eqn:zzbarPoincare} is in conformal gauge, as is custom in string theory. However, the Riemann surfaces considered in string theory are compact and have no analogue of the singularity at $|z|=1$. For example, the sphere $S^2$ has a conformal factor $g_{z{\bar z}}= 2(1+|z|^2)^{-2}$ that is well-behaved for any $|z|$. It is because of this crucial difference that AdS$_2$ can have infinitely many QHD's, while $S^2$ has none. Conversely, AdS$_2$ has no conformal Killing vectors (CKV's), the normalizable vector fields that leave the metric invariant even after fixing to conformal gauge, while $S^2$ has 3. 

It is also instructive to compare with non-critical string theory (and its closely related matrix model descriptions). Again, it is customary to fix conformal gauge at the outset, but the conformal (Weyl) factor becomes a dynamical degree of freedom which, it turns out, is described by Liouville theory \cite{Distler:1988jt,Engelsoy:2016xyb,Mertens:2017mtv}. In contrast, in our construction we insist that the conformal factor is fixed (and singular), and this forces coordinate transformations to act only on the QHD 
\eqref{eqn:qhd} which, accordingly, parametrizes the space of states.

\section{The Effective Theory of the Near-Horizon Geometry }
\label{sec:effthy}

The diffeomorphisms \eqref{eqn:inftrans} relate vacua by a transformation that is a symmetry, except that it acts on a space of distinct boundary conditions at $|z|=1$. In this situation low-lying states can be realized by ``moving slowly" among all these configurations. Nonlinear realization of symmetry offers a systematic formalism that implements this intuition. 

In order to proceed, the algebra of infinitesimal symmetries \eqref{eqn:inftrans} must be exponentiated to map out the corresponding group manifold.
At first this seems difficult to carry out, because the algebra has infinitely many generators, but the problem here has special properties that help out. A finite diffeomorphism transforms the QHD \eqref{eqn:qhd} as a tensor, except for an anomalous contribution with the infinitesimal form \eqref{eqn:hzzinf}. The finite version of the latter is determined by the requirement that successive coordinate transformations must compose appropriately, {\it ie$.$} consistently with the Leibniz rule. This condition is familiar from conformal field theory in two dimensions where the finite form of (projectively realized) conformal symmetry is given by the Schwarzian. We can adapt this result to represent the general QHD \eqref{eqn:qhd} as a finite diffeomorphism $f(z)$ that 
introduces 
\begin{equation}
\label{eqn:hzzinf2}
h_{zz} dz^2 =   -2 \left( \frac{\partial^3_z f(z)}{\partial_z f(z)}   - \frac{3}{2}  \left(\frac{\partial^2_z f(z)}{\partial_z f(z)}\right)^2   \right)  dz^2  ~.
\end{equation}
This formula expresses the entire space of non-trivial spacetime geometries in a closed form.\footnote{The generator of diffeomorphisms \eqref{eqn:inftrans} is ``exact" in that there is no restriction on $|z|$ (but it is ``infinitesimal") while the exponentiation \eqref{eqn:hzzinf2} is ``exact" in that the transformation is finite (but it is at asymptotic order $(1-|z|^2)^2$). The underlying truly ``exact" mathematical structure is the universal Teichm\"{u}ller space.} 

We have stressed the representation of the group theory in terms of two-dimensional spacetimes, as befits the interpretation in term of near-horizon black hole geometry. However, low energy effective quantum field theory derives its power by exploiting the principles of symmetry, without appealing to a particular realization. We can think of the non-normalizable diffeomorphisms \eqref{eqn:xin} as abstract generators that close on themselves and so form a symmetry algebra. More precisely, they close on an augmented set that includes the vectors $\epsilon(z) = 1, z, z^2$. However, in the geometrical realization these ``additional" modes do not change the metric at all, they generate $SL(2)$, the isometry group of AdS$_2$. Therefore, upon exponentiation the generators of the symmetry algebra do not quite form a group, because of the caveat they define a coset. The algebra identifies the coset generated by the non-normalizable diffeomorphisms as ${\rm Diff(S^1)}/ SL(2)$. 

The group ${\rm Diff}(S^1)$ is defined in terms of its action on the circle. There is a canonical map from the QHD's \eqref{eqn:qhd} parametrizing deformations of AdS$_2$ to the periodic functions of a single real variable $\tau$ on the circle $S^1$ ``at infinity" via the identifications $z=e^{i\tau}$ and ${\bar z}=e^{-i\tau}$ . This map is subtle in that the boundary $|z|=1$ does not belong to the AdS$_2$ geometry which is non-compact because its conformal factor diverges as $|z|\to 1$. However, it makes sense anyway because the QHD's \eqref{eqn:hzzinf} have no intrinsic scale, they depend only on the complex structure which is perfectly smooth as $z\to 1$. This is the natural generalization to AdS$_2$/CFT$_1$ of the holographic map that applies to the AdS/CFT correspondence in higher dimensions. 

Representations of ${\rm Diff}(S^1)$ have been studied in great detail by mathematicians (for a review see \cite{Oblak:2016eij}). Proper representations simply keep the length of the circle invariant but, in quantum mechanics, it is sufficient to consider projective representations where the group elements transform state vectors among themselves only up to a phase, as long as the phase is assigned consistently with the group structure. An important aspect of the mathematical classification of projective representations is the stabilizer group, comprised of the group elements that act trivially in the representation at hand. The stabilizer group of most projective representations includes $U(1)$, corresponding to the translations along $S^1$, but only in the case of the so-called ``identity" representation is this $U(1)$ enhanced to the $SL(2)$ subgroup of ${\rm Diff}(S^1)$. Moreover, in this particular case the group action is proportional to \eqref{eqn:hzzinf}, with the understanding that the complex number is a pure phase $z = e^{i\tau}$. Therefore, from the purely formal point of view, there is a unique projective representation of the coset ${\rm Diff}(S^1)/SL(2)$ that is viable for the description of AdS$_2$ at low energy \cite{Stanford:2017thb}. The fact that it coincides with the 
one identified by an explicit analysis of the near-horizon geometry of the black hole illustrates the power of low energy effective field theory.

In general spacetime dimension, the energy-momentum tensor in quantum field theory is related to the action as
\begin{equation}
T^{\mu\nu} = -  \frac{2\pi}{\sqrt{g}} \frac{\delta I}{\delta g_{\mu\nu}}~,
\label{eqn:emtensor}
\end{equation}
and diffeomorphism invariance ensures that $T^{\mu\nu}$ is conserved. Presently, the time $\tau$ is the only dimension in the boundary theory so there is just one component, the energy. Moreover, it is only conserved if it is a constant. Therefore, the formula  \eqref{eqn:emtensor} for the energy-momentum ``tensor" --- it is a single number in one (space)time dimension --- shows that the action must be proportional to the metric coefficient in
\eqref{eqn:hzzinf2} and so determines the effective Lagrangian
\begin{equation}
I =  \frac{C_T}{T} \int \frac{d\tau}{2\pi} 
\left[ \frac{1}{2} (\partial_\tau f)^2 + \frac{\partial^3_\tau f}{\partial_\tau f}  - \frac{3}{2}  \left(\frac{\partial^2_\tau f}{\partial_\tau f}\right)^2  
 \right]~.
 \label{eqn:schct}
\end{equation}
The overall constant of proportionality $C_T/T$ is arbitrary for now, to be discussed further in the subsequent section. 

All black hole states, corresponding to a general finite diffeomorphism $f(\tau)$, is assigned the same energy, so we interpret them as degenerate vacua. We pick the specific reference geometry \eqref{eqn:zzbarPoincare} as ``the" vacuum because it ``looks" simple but, as usual for a 
spontaneously broken symmetry, any other reference would be physically equivalent. The invariant physical reality is that the (non-normalizable) symmetry generators act on the (arbitrarily chosen) reference vacuum and create physical states that are Goldstone bosons of the broken symmetry. 

At the risk of beating a dead horse, we stress that there are (at least) two distinct perspectives on the computation leading to the Schwarzian 
action \eqref{eqn:schct}:
\begin{itemlist}
\item
It is {\it quantum gravity} in two dimensions: a theory of excitations in the near-horizon region of black hole. 

\item
It is an effective theory of symmetries that act entirely on the asymptotic boundary $S^1$. This is a scale invariant theory {\it without gravity}.
\end{itemlist}
The equivalence of these two descriptions is the content of the modern realization of AdS$_2$/CFT$_1$ holography. The presentation here (and elsewhere) 
{\it proves} this duality at the level of effective field theory. 

That the arguments motivating (and largely establishing) AdS$_{d+1}$/CFT$_{d}$ holography in dimensions $d>1$ fail in $d=1$ were recognized already in \cite{Maldacena:1997re}
and was frequently interpreted as AdS$_2$/CFT$_1$ holography somehow not being possible \cite{Maldacena:1998u,Spradlin:1999bn,Maldacena:1998u,Castro:2008ms,Castro:2014ima}. The modern understanding developed over the last few years is that much 
the same arguments apply, after all, but at the level of effective quantum field theory. Thus nearAdS$_2$/nearCFT$_1$ holography is stronger in that it is proven, but it is weaker in that it just identifies IR theories, no full duality is claimed.

\section{The Symmetry Breaking Scales}
\label{sec:ssb}

The derivation of the Schwarzian \eqref{eqn:schct} does not determine its overall coefficient, the dimensionful symmetry breaking scale $C_T/T$. In low energy effective theory it is an arbitrary parameter but, in the context of a specific UV completion, such as the SYK model, it can be derived from microscopic parameters \cite{Maldacena:2016hyu,Bagrets:2016cdf}. 
Alternatively, it can be related to the parameters of a particularly simple 2D gravity model, such as JT-gravity \cite{Maldacena:2016upp}. In the application to black holes, the scale is determined by the complete black hole geometry, going beyond the near-horizon limit. In this setting there is an appealing geometrical interpretation of the symmetry breaking scale. 

The near-horizon geometry \eqref{eqn:zzbarPoincare} has a single 
overall scale, the AdS$_2$ curvature $\ell_2=Q$ (which we mostly suppress to unclutter formulae), but its metric diverges at $|z|=1$ where the ``throat" attaches to surrounding geometry. The full (not ``near-horizon") 
geometry extends farther, to the region where $|z|>1$. There the area of the transverse sphere $S^2$ is slightly larger, and the larger radius of the Euclidean circle
implies a non-zero local temperature. The matching with the UV scale amounts to the ratio of these two features, it evaluates the radial derivative of the area element 
at $|z|=1$. Thus the black hole interpretation captures the UV data by extending the geometry into the region $|z|>1$, but only infinitesimally. In contrast, the low energy effective 
theory, as a theory in itself, regulates the divergence at $|z|=1$ and focuses on the interior $|z|<1$. 

To make these principles explicit, consider a spherically symmetric black hole in $D=4$ spacetime dimensions, presented in Schwarzschild-like coordinates as
\[
ds^2 = - g(r) dt^2 + g^{-1}(r) dr^2 + r^2 d\Omega^2_2~.
\]
The function $g(r)$ depends only on the radial coordinate $r$, as indicated by its argument. 
It vanishes at the horizon $\left.g\right|_{r=r_{\rm hor}} =0$ and its derivative there gives the black hole temperature $T = \left.(4\pi)^{-1} \partial_r g \right|_{r=r_{\rm hor}}$.  
In the context of this presentation, we focus on the Reissner-Nordstr\"{o}m black holes where
\[
	g(r)   =  1  -  \frac{2M}{r} + \frac{Q^2}{r^2}~,
\]and especially the supersymmetric limit where $Q=M$ where $g_{\rm BPS} = (1 - Q/r)^2$. In this case the derivative $\partial_r g $ vanishes at the horizon $r_{\rm hor} = Q$ so the temperature vanishes in the BPS limit, as expected. 

A straightforward way to analyze near-extremal black holes is to relax the BPS condition $Q=M$ from the outset and then study the limit as $M\to Q_+$. The alternative procedure we pursue, as outlined above, is conceptually distinct and, especially when rotation is incorporated, much simpler computationally. Starting from the strictly supersymmetric black hole, we simply evaluate the temperature $T = (4\pi)^{-1} \partial_r g$ and the entropy $S = (4G_4)^{-1}4\pi r^2$ just {\it outside} the horizon: 
\begin{equation}
\frac{C_T}{T} = \frac{\Delta S}{\Delta T} =  \frac{4\pi }{\Delta \left(\partial_r g\right)} ~ \Delta \pi r^2 = \frac{8\pi^2 r}{\partial^2_r g}~,
\label{eqn:betterCT}
\end{equation}
with the final expression evaluated at the horizon. The BPS function has second derivative $g''_{\rm BPS} = 2/Q^2$ at the horizon and this value returns the correct symmetry breaking scale \eqref{eqn:SSBscale}. The equivalence of moving away from the horizon (in a given BPS solution) and deforming the solution away from the BPS limit (but always considering the event horizon) was recognized and developed in \cite{Larsen:2018iou,Hong:2019tsx}. 

The entire geometry of a BPS black hole preserves supersymmetry, but it is enhanced to superconformal symmetry in the near-horizon region. The construction presented here relies only on the BPS geometry, so it computes the scale associated with the breaking of conformal symmetry without breaking supersymmetry at the same time. Therefore, we expect that the scale associated with the breaking of conformal symmetry enjoys protection by supersymmetry. 

There is another aspect of supersymmetry that we have, for simplicity, suppressed in the last several sections: we focus on the gravity sector, as expected for a discussion of black holes, but the Einstein-{\it Maxwell} theory also features a gauge field. Its symmetry structure is analogous to the one discussed for diffeomorphisms, but much simpler. There are normalizable gauge fields in the AdS$_2$ region: 
\[
A_z= \partial_z \sigma(z) = z^{n-1} ~, ~~~~ n=1, 2, \ldots. 
\]
They are formally ``pure" gauge but the applicable gauge function $\sigma(z)$ is not normalizable in the background metric. Therefore, these modes are in fact physical. The effective action for the
$U(1)$ field is a simple free scalar. Combining these modes from the gauge sector with the gravitational modes \eqref{eqn:betterCT}, the effective low energy theory of near-extremal Kerr-Newman black holes becomes
\begin{equation}
\label{eqn:N=2act}
I =  \frac{C_T}{T} \int \frac{d\tau}{2\pi} 
\left[ \frac{1}{2}(\partial_\tau f)^2 + \frac{\partial^3_\tau f}{\partial_\tau f}  - \frac{3}{2}  \left(\frac{\partial^2_\tau f}{\partial_\tau f}\right)^2  
- \frac{1}{2}(\partial_\tau\sigma)^2 \right]~.
\end{equation}
As before, this effective action describes radial excitations of the near-extremal black hole in terms of an ``angular" variable, the Euclidean time $\tau$. However, it also incorporates 
a more conventional angular feature: a rotating black hole mixes the azimuthal angle of the $S^2$ horizon, that we have suppressed so far, with ``the" time. 
The reference to $\tau$ as a ``coordinate" and  $\sigma$  as a ``gauge parameter" emphasizes our focus on 2D AdS$_2$ (and its 1D boundary $S^1$) over its 4D 
progenitor AdS$_2\times S^2$, not to mention the supersymmetric extensions of these geometries.  

This difference in perspectives notwithstanding, we have taken care that $\tau$ and $\sigma$ both have canonical periodicity $2\pi$. As usual, fermionic wave ``functions" are 
more precisely sections of spin bundles and, as such, they can be antiperiodic. However, this caveat must be consistently applied, in that fermions that are antiperiodic around the 
Euclidean time $\tau$ must also be antiperiodic around the gauge ``coordinate" $\sigma$. It is in this precise sense that the two terms in the action \eqref{eqn:N=2act} have the ``same" normalization. In the version of the setup that incorporates gravitini, the identity of these coefficients is also required by ${\cal N}=2$ supersymmetry in one (space)time dimension \cite{Fu:2016vas}.

The generalized Schwarzian \eqref{eqn:N=2act} encodes the basic thermodynamics of the low-lying excitations above the supersymmetric ground state, as well as much else.
The linear functions $f = \tau T$, and $\sigma = \Omega\tau/2\pi$ are well-defined (with coordinate periodicity $\Delta\tau = 2\pi\beta$) 
and together give the mass formula 
\[
M =  Q ~~ +  ~~ {1\over 2} \left(\frac{C_T}{T}\right) ~ \left[ T^2  + \left( \frac{\Omega}{2\pi}\right)^2 \right]~.
\]
As we have stressed previously, the $T^2$ dependence makes a statement about scaling symmetry. The overall coefficient is arbitrary in the low energy theory, but determined by 
matching with the UV theory. At this point we stress that the {\it relative} coefficient between the two terms is determined by ${\cal N} = 2$ supersymmetry. 
At the classical level, this claim applies even for Einstein-Maxwell theory, with no gravitini in the theory, but generally the two conceptually distinct response coefficients may differ at the quantum level.

The interrelation between radial/temporal geometry AdS$_2$ and the horizon $S^2$ can be illuminated by comparison with asymptotically flat black holes in 5D which, in the circumstances that are best developed, can be understood in terms of limits of a near-horizon AdS$_3\times S^3$ geometry. In this context the isometries of the AdS$_3$ and $S^3$ factors are $SL(2)^2$ and $SU(2)^2$, respectively, and rotation is interpreted geometrically through coordinate identifications that fibrate an $SU(2)$ nontrivially over an $SL(2)$ factor \cite{Cvetic:1998xh}. 
In the dual CFT$_2$ with ${\cal N}=(4,4)$ supersymmetry rotation is implemented by the spectral flow automorphism. At the classical level, these interrelations ensure
that the levels of the underlying $SL(2)$ and $SU(2)$ current algebras coincide, a precise analogue of the agreement between the prefactors of $T^2$ and $\left( \Omega/2\pi\right)^2$ 
in the mass formula. Generally, there are nontrivial corrections in the quantum theory and it is interesting that, in the AdS$_3\times S^3$ setup, they may be determined by anomalies \cite{Kraus:2005vz,Kraus:2005zm}. 

Some instances of AdS$_2\times S^2$ can be realized as limits of AdS$_3\times S^3$ 
\cite{Gupta:2008ki,Balasubramanian:2009bg,Castro:2010vi}, but others cannot, and such limits may or may not be consistent with supersymmetry. 
It is interesting to inquire how quantum corrections to the linear response coefficient $C_T/T$ arise directly in four spacetime dimensions and how such precision information may 
descend to the simple quantum mechanics of horizon modes we consider from genuine quantum field theory in higher dimensions.

\section{Quantum Corrections to Black Hole Entropy}
\label{sec:quantum}

There is a well-defined notion of logarithmic quantum corrections to the Bekenstein-Hawking area law:
\begin{equation}
S = \frac{A}{4G_4}  -  2  a  \log A + \cdots~,
\label{eqn:logA}
\end{equation}
when all macroscopic lengths are scaled uniformly \cite{Sen:2012kpz}. 
The correction is due to virtual loops of massless fields in the near-horizon region of the black hole. Therefore, the coefficient $a$ can be computed unambiguously 
in effective quantum field theory and doing so adds perspective to the power and limitations of the effective field theory description of black holes.\footnote{We choose the notation $a$ for the coefficient because, in many $D=4$ examples, there is a close relation to the corresponding Weyl anomaly coefficient. It is unfortunate that, with this convention, each unit anomalous mass dimension of the entropy 
$M\partial_M \Delta S = + 1 $  contributes $\Delta a = 1/4$. }

The explicit computations of quantum corrections to black holes can be quite involved. The most laborious part is to compute functional determinants for fields that are non-vanishing in the classical black hole solution. Other fields must be taken into account as well, because they also contribute to virtual loops, and such spectator fields may couple nontrivially to the background. 
More conceptually, zero-modes of the entire black hole must be treated correctly and, at the quantum level, statistical ensembles are not all equivalent so care must be taken that the entropy is microcanonical (the energy is fixed but the temperature is not). These various features were known for a long time, 
but they were greatly developed and refined about a decade ago, especially by A. Sen 
\cite{Banerjee:2010qc,Bhattacharyya:2012wz,Keeler:2014bra,Charles:2015eha}. A highlight of this advance was the case of black holes in string theory (that preserve a sufficient amount of supersymmetry) where the coefficients from loops of massless particles computed in effective field theory were shown to agree precisely with the corresponding expansion of microscopic counting formulae \cite{Banerjee:2011jp,Sen:2014aja}. This result gives confidence that the various contributions to the logarithmic corrections have been correctly understood. 

To make this general discussion more concrete, we now return to the $D=4$ Kerr-Newman black hole and specifically its ground state entropy: 
\begin{equation}
S = \frac{A_0}{4G_4}  -  2 a_0 \log A_0  + \cdots   ~,
\label{eqn:logA0}
\end{equation}
where $A_0 = 4\pi Q^2$. As we emphasized in the preceding paragraph, there are several comparable contributions to the coefficient $a_0$. It is interesting to compare them to the description by Schwarzian action \eqref{eqn:schct}.
This is possible because unknown physics at very high energy, depending on the Planck scale, string scale, or possible compactification scales, all decouple. It contributes to local operators suppressed by a power of energy, rather than a logarithm. 
Thus the relevant physics is at low energy, yet care must be taken because, as in most effective descriptions, the range where the Schwarzian description applies is limited at both high and low energy. 

In the spherical reduction of Einstein-Maxwell theory the propagating modes of gravity and the gauge field have spherical harmonics $l\geq 2$ and $l\geq 1$, respectively. The scale of all these modes is set by $Q$, the radius of $S^2$, 
so they are too heavy to be dynamical degrees of freedom at the much longer length scale of the Schwarzian theory $Q^3$. However, the couplings in an effective field theory are determined at the highest energies of its applicability, by matching with the UV theory, so the Kaluza-Klein modes can leave an imprint at low energy. We can interpret such contributions as running of the effective low energy cosmological constant due to virtual effects. Numerically, gravity and vector fields couple to one another and combine to a contribution $\Delta a_0 = 53/45$. The completion to ${\cal N}=2$ supergravity by gravitini yields an additional $\Delta a_0 = - 589/360$ for a total of $\Delta a_0=-11/24$ from all Kaluza-Klein modes. The precise numerical coefficients are unimportant in this presentation, we merely note that they are computable and of order one. 

The position and orientation of the black hole break translational and rotational symmetry, respectively. These collective degrees of freedom can produce Goldstone modes that remain in the low energy path integral and then their quantum fluctuations contribute with a weight determined by the scaling limit defining \eqref{eqn:logA}. At strictly zero temperature and nonvanishing rotation the integral over $3$ translational zero-modes contribute $\Delta a_0 =  3 \times (1/4) = 3/4$ \footnote{The factor of $1/4$ was explained in the previous footnote.}. In this case the corresponding addition $-(3/2)\log A = - 3 \log ({\rm length})$ to the entropy is simply due to a single factor of the three-dimensional spatial volume in the partition function.\footnote{The dependence of the zero-mode contribution on the theory and the ensemble is far from trivial. It was discussed by Sen in many interesting situations \cite{Sen:2008vm,Sen:2012dw,Sen:2012kpz,Sen:2014aja}. The results were summarized in an Appendix of \cite{Charles:2015eha}. For reference, the completion to a spherically symmetric black hole in ${\cal N}=2$ supergravity gives a total contribution $\Delta a_0 = -1/2$ from zero-modes.}

In our description, we can understand the dependence on the translational mode as a quantum anomaly. If we expand the quantum geometry as
\[
g_{zz} = \sum_n c_n g_n~,
\]
where the modes
\[
g_n = \sqrt{\frac{(n^2-1)n}{2\pi \ell^2_2}}z^{n-2}~,
\]
are normalized so $\int  d^2 z \sqrt{g} |g_n|^2 = 1$ on AdS$_2$, the path integral measure 
\[
{\cal D}\phi = \prod_{n=2}^\infty dc_n~,
\]
clearly depends on the AdS$_2$ scale $\ell_2$. This quantum dependence introduces the anomalous scaling 
$- 3 \log ({\rm length})$
\[
\ell_2 \frac{\partial}{\partial\ell_2} \ln {\cal D}\phi = - 2\pi\ell^2_2 \sum_{n=2}^\infty |g_n |^2 =  - 2\pi\ell^2_2 (1-|z|^2)^4  \sum_{n=2}^\infty \frac{(n^2-1)n}{2\pi \ell^2_2}|z|^{2(n-2)}=  - 3~.
\]
The first equality assigned the usual value 
\[
\int \sqrt{g} d^2 z = \int_0 \frac{2\ell^2_2}{(1-|z|^2)^2}  \cdot 2\pi \cdot \frac{1}{2} d|z|^2 = - 2\pi\ell^2_2~,
\] 
as the renormalized volume of AdS$_2$. At high temperature this quantum dependence on the zero-mode precisely cancels the volume dependence $\sim V_3T^3$ of the surrounding thermal gas \cite{Sen:2012dw}. 

The quantum correction to the excitations described by the Schwarzian is due to the very same modes as the volume dependence of the vacuum. However, because of the boundary condition that identifies $z$ and ${\bar z}$, it captures only $1/2$ of the anomalous dimension: 
\[
\frac{\partial}{\partial\log({\rm length})} \ln Z = -\frac{3}{2}~.
\]
The anomalous scaling $\log Z_{\rm Sch} \sim (3/2)\log T$ for the thermal partition function of the Schwarzian model was computed by 
many researchers  \cite{Maldacena:2016hyu,Stanford:2017thb,Charles:2019tiu,Heydeman:2020hhw}. 
It can be understood as the ``missing" $SL(2)$ volume that is removed in the ${\rm Diff}(S^1)/SL(2)$ coset. Our computation relates this anomalous dimension to explicit physical modes. 

The agreement between scaling dimensions hides an important distinction between scales. To explain, we return to the generic form of the logarithmic quantum correction, presented in \eqref{eqn:logA} as dependence on black hole parameters through $\log A$, with the understanding that the dimensionful area $A$ must be evaluated in units that are kept fixed as physical characteristics of the black hole are rescaled. The implicit reference scale that sets units could be the Planck length $\sim G^{1/2}_4$, because it is the same for all black holes. 
However, according to Wilson's physical picture of renormalization it is more appropriate to apply a reference scale set by the fluctuations themselves. For quantum corrections to the supersymmetric ground state we can pick a ``standard" black hole with scale $Q_0$ as a reference and then estimate quantum corrections to other black holes in the family as $\log Q^2/Q^2_0$ with some coefficient we compute. 

For logarithmic corrections computed by the Schwarzian theory we have in mind a different situation. Given a single reference black hole with scale $Q_0$ we study different temperatures $T$ and find logarithmic corrections $\log T$. In this setup the typical length scale of the thermal fluctuations $C_T/T\sim Q^3_0/G_4$ is much larger than 
the AdS$_2$ scale $\ell_2\sim Q_0$. Thus the Kaluza-Klein modes make their imprint on the ground state entropy at energies that are much higher than where 
the Schwarzian description applies. 

However, at {\it very} low temperatures $T \ll Q^3$ the Schwarzian theory fails because the energies determined by the AdS$_2$ description becomes unreliable, the spectrum incorporates 
wave lengths that probe beyond the ``throat" with scale $\sim Q^3$.  The low energy spectrum of an asymptotically flat black hole in equilibrium with its surroundings has no lower limit, there is no gap, so modes at extremely low temperature ultimately dominate the entropy of a black hole in flat space because they contribute $S\sim V_3T^3$ where the volume $V_3$ can be taken arbitrarily large. Such modes obviously do not characterize the black hole, they are features of the surrounding flat space, and so they should not be included in the accounting of the black hole entropy. Mathematically, the path integral with black hole boundary conditions must be modulo a factor that diverges as the spatial volume $V_3\to\infty$ with $T$ fixed \cite{Sen:2012dw}. 

Technical as this point may be, it is important to note that the volume $V_3$ introduces another scale, an IR regulator. This is significant even for strictly supersymmetric black holes where the scale regulating the contribution of KK-modes to the logarithmic correction \eqref{eqn:logA0} is the horizon scale $\ell_2$, while the appropriate benchmark for the translational zero-modes is the IR volume $V_3$. This feature allows a smooth transition between the quantum correction \eqref{eqn:logA0} and the quantum correction to the Schwarzian description in the low (but finite) temperature limit.

\section{Summary}
\label{sec:summ}

This presentation discussed black hole thermodynamics with emphasis on near-extremal black holes. This low-temperature regime is described by the Schwarzian 
theory with Lagrangian \eqref{eqn:schct}. 
In the literature, there has been great interest in this theory, as an IR limit of the SYK-model \cite{Sachdev:1992fk,Sachdev:2010um,Kitaev,
	Jevicki:2016bwu,Maldacena:2016hyu,Jevicki:2016ito,Fu:2016vas,Turiaci:2017zwd,Stanford:2017thb,Liu:2019niv,Maldacena:2019ufo},
and as the low energy effective description of JT-gravity
\cite{Teitelboim:1983ux,Jackiw:1984je,Almheiri:2014cka,Jensen:2016pah,Maldacena:2016upp,Engelsoy:2016xyb,Forste:2017apw,Forste:2017kwy,
Cvetic:2016eiv,Nayak:2018qej,Moitra:2018jqs,Saad:2019lba,Iliesiu:2019xuh,
Mertens:2019tcm,Stanford:2019vob,Iliesiu:2019lfc,Iliesiu:2020qvm}.
We seek to complement these mainstream points of view by further developing the connection to spacetime black holes. 

In this spirit, our main result is to present the explicit wave functions in two spacetime dimensions that dominate the black hole thermodynamics at low temperature. 
The modes we focus on are zero-modes in the two-dimensional spacetime, in the sense that they can be generated by diffeomorphisms. 
However, because the required coordinate transformations are not normalizable, these modes are in fact physical and, it turns out, described via the Schwarzian 
action. 

The special modes in AdS$_2$ that we highlight appear prominently in discussions of logarithmic corrections to supersymmetric black hole entropy. 
In our interpretation these modes can be identified with the translational zero-modes that become normalizable at finite temperature. We support this identification by showing that
the contribution to the black hole entropy from zero-modes to the entropy of strictly supersymmetric black holes agrees with the corresponding limit of black holes at finite temperature. 
This success gives confidence that the modes we focus on dominate at low energy. 

The study of black holes at very low temperature is promising for illuminating major issues of current interest, such as the black hole information loss paradox (some recent reviews \cite{Almheiri:2020cfm,Raju:2020smc}) and its relation to quantum chaos \cite{Cotler:2016fpe,Saad:2018bqo}. 
Some studies in these research directions focus on specific UV-completions, such as the SYK-model, that depend on many additional details. Others consider properties of the
Euclidean quantum path integral which, by its nature, is not constructive. 

Our results highlight that the supersymmetric ground state is largely inert, only a limited number of modes contribute to the entropy, and the UV aspects of the theory are decoupled as well. 
The explicit spacetime wave functions also permit the couplings to the ambient spacetime, beyond the near-horizon region, to be determined precisely. 
Therefore, this simplified setting may offer a useful toy model that captures the essential parts of the quantum information flow between the black hole and its environment. 

\section*{Acknowledgements}

This research was supported in part by the U.S. Department of Energy under grant DE-SC0007859.
SC is supported by the Samsung Scholarship, the Leinweber Graduate Fellowship and the Rackham Predoctoral Fellowship. We thank Luca Iliesiu and Joaquin Turiaci for comments.

\bibliographystyle{ws-rv-van}
\bibliography{references}

\begin{thebibliography}{72}
\providecommand{\natexlab}[1]{#1}
\providecommand{\url}[1]{\texttt{#1}}
\expandafter\ifx\csname urlstyle\endcsname\relax
  \providecommand{\doi}[1]{doi: #1}\else
  \providecommand{\doi}{doi: \begingroup \urlstyle{rm}\Url}\fi

\bibitem{Sachdev:1992fk}
S.~Sachdev and J.~Ye, {Gapless spin fluid ground state in a random, quantum
  Heisenberg magnet}, \emph{Phys. Rev. Lett.} {\bf 70}, \penalty0 3339,
  (1993).
\newblock \doi{10.1103/PhysRevLett.70.3339}.

\bibitem{Sachdev:2010um}
S.~Sachdev, {Holographic metals and the fractionalized Fermi liquid},
  \emph{Phys. Rev. Lett.} {\bf 105}, \penalty0 151602,  (2010).
\newblock \doi{10.1103/PhysRevLett.105.151602}.

\bibitem{Kitaev}
A.~Kitaev, A simple model of quantum holography.  (Feb. 12, April 7, and May
  27, 2015).
\newblock URL \url{http://online.kitp.ucsb.edu/online/entangled15/}.

\bibitem{Rosenhaus:2018dtp}
V.~Rosenhaus, {An introduction to the SYK model}, \emph{J. Phys. A}. {\bf 52},
  \penalty0 323001,  (2019).
\newblock \doi{10.1088/1751-8121/ab2ce1}.

\bibitem{Maldacena:2016upp}
J.~Maldacena, D.~Stanford, and Z.~Yang, {Conformal symmetry and its breaking in
  two dimensional Nearly Anti-de-Sitter space}, \emph{PTEP}. {\bf
  2016}\penalty0 (12), \penalty0 12C104,  (2016).
\newblock \doi{10.1093/ptep/ptw124}.

\bibitem{Larsen:2018iou}
F.~Larsen, {A nAttractor mechanism for nAdS$_{2}$/nCFT$_{1}$ holography},
  \emph{JHEP}. {\bf 04}, \penalty0 055,  (2019).
\newblock \doi{10.1007/JHEP04(2019)055}.

\bibitem{Fu:2016vas}
W.~Fu, D.~Gaiotto, J.~Maldacena, and S.~Sachdev, {Supersymmetric
  Sachdev-Ye-Kitaev models}, \emph{Phys. Rev. D}. {\bf 95}\penalty0 (2),
  \penalty0 026009,  (2017).
\newblock \doi{10.1103/PhysRevD.95.026009}.
\newblock [Addendum: Phys.Rev.D 95, 069904 (2017)].

\bibitem{Forste:2017apw}
S.~F\"orste, J.~Kames-King, and M.~Wiesner, {Towards the Holographic Dual of N
  = 2 SYK}, \emph{JHEP}. {\bf 03}, \penalty0 028,  (2018).
\newblock \doi{10.1007/JHEP03(2018)028}.

\bibitem{Coleman:1991ku}
S.~R. Coleman, J.~Preskill, and F.~Wilczek, {Quantum hair on black holes},
  \emph{Nucl. Phys. B}. {\bf 378}, \penalty0 175--246,  (1992).
\newblock \doi{10.1016/0550-3213(92)90008-Y}.

\bibitem{Kim:2006he}
S.~Kim and K.-M. Lee, {1/16-BPS Black Holes and Giant Gravitons in the AdS(5) X
  S**5 Space}, \emph{JHEP}. {\bf 12}, \penalty0 077,  (2006).
\newblock \doi{10.1088/1126-6708/2006/12/077}.

\bibitem{Hosseini:2017mds}
S.~M. Hosseini, K.~Hristov, and A.~Zaffaroni, {An extremization principle for
  the entropy of rotating BPS black holes in AdS$_{5}$}, \emph{JHEP}. {\bf 07},
  \penalty0 106,  (2017).
\newblock \doi{10.1007/JHEP07(2017)106}.

\bibitem{Choi:2018hmj}
S.~Choi, J.~Kim, S.~Kim, and J.~Nahmgoong, {Large AdS black holes from QFT}
  (10. 2018).

\bibitem{Preskill:1991tb}
J.~Preskill, P.~Schwarz, A.~D. Shapere, S.~Trivedi, and F.~Wilczek,
  {Limitations on the statistical description of black holes}, \emph{Mod. Phys.
  Lett.} {\bf A6}, \penalty0 2353--2362,  (1991).
\newblock \doi{10.1142/S0217732391002773}.

\bibitem{Maldacena:1996ds}
J.~M. Maldacena and L.~Susskind, {D-branes and fat black holes}, \emph{Nucl.
  Phys. B}. {\bf 475}, \penalty0 679--690,  (1996).
\newblock \doi{10.1016/0550-3213(96)00323-9}.

\bibitem{Cvetic:1998xh}
M.~Cvetic and F.~Larsen, {Near horizon geometry of rotating black holes in
  five-dimensions}, \emph{Nucl. Phys. B}. {\bf 531}, \penalty0 239--255,
  (1998).
\newblock \doi{10.1016/S0550-3213(98)00604-X}.

\bibitem{Dijkgraaf:2000fq}
R.~Dijkgraaf, J.~M. Maldacena, G.~W. Moore, and E.~P. Verlinde, {A Black hole
  Farey tail} (5. 2000).

\bibitem{Kraus:2006nb}
P.~Kraus and F.~Larsen, {Partition functions and elliptic genera from
  supergravity}, \emph{JHEP}. {\bf 01}, \penalty0 002,  (2007).
\newblock \doi{10.1088/1126-6708/2007/01/002}.

\bibitem{Kunduri:2007vf}
H.~K. Kunduri, J.~Lucietti, and H.~S. Reall, {Near-horizon symmetries of
  extremal black holes}, \emph{Class. Quant. Grav.} {\bf 24}, \penalty0
  4169--4190,  (2007).
\newblock \doi{10.1088/0264-9381/24/16/012}.

\bibitem{Camporesi:1994ga}
R.~Camporesi and A.~Higuchi, {Spectral functions and zeta functions in
  hyperbolic spaces}, \emph{J. Math. Phys.} {\bf 35}, \penalty0 4217--4246,
  (1994).
\newblock \doi{10.1063/1.530850}.

\bibitem{Banerjee:2010qc}
S.~Banerjee, R.~K. Gupta, and A.~Sen, {Logarithmic Corrections to Extremal
  Black Hole Entropy from Quantum Entropy Function}, \emph{JHEP}. {\bf 03},
  \penalty0 147,  (2011).
\newblock \doi{10.1007/JHEP03(2011)147}.

\bibitem{Sen:2012kpz}
A.~Sen, {Logarithmic Corrections to N=2 Black Hole Entropy: An Infrared Window
  into the Microstates}, \emph{Gen. Rel. Grav.} {\bf 44}\penalty0 (5),
  \penalty0 1207--1266,  (2012).
\newblock \doi{10.1007/s10714-012-1336-5}.

\bibitem{Larsen:2014bqa}
F.~Larsen and P.~Lisbao, {Quantum Corrections to Supergravity on AdS$_2\times
  S^2$}, \emph{Phys. Rev. D}. {\bf 91}\penalty0 (8), \penalty0 084056,  (2015).
\newblock \doi{10.1103/PhysRevD.91.084056}.

\bibitem{Charles:2019tiu}
A.~M. Charles and F.~Larsen, {A one-loop test of the
  near-AdS$_{2}$/near-CFT$_{1}$ correspondence}, \emph{JHEP}. {\bf 07}\penalty0
  (07), \penalty0 186,  (2020).
\newblock \doi{10.1007/JHEP07(2020)186}.

\bibitem{Distler:1988jt}
J.~Distler and H.~Kawai, {Conformal Field Theory and 2D Quantum Gravity},
  \emph{Nucl. Phys. B}. {\bf 321}, \penalty0 509--527,  (1989).
\newblock \doi{10.1016/0550-3213(89)90354-4}.

\bibitem{Engelsoy:2016xyb}
J.~Engels\"oy, T.~G. Mertens, and H.~Verlinde, {An investigation of AdS$_{2}$
  backreaction and holography}, \emph{JHEP}. {\bf 07}, \penalty0 139,  (2016).
\newblock \doi{10.1007/JHEP07(2016)139}.

\bibitem{Mertens:2017mtv}
T.~G. Mertens, G.~J. Turiaci, and H.~L. Verlinde, {Solving the Schwarzian via
  the Conformal Bootstrap}, \emph{JHEP}. {\bf 08}, \penalty0 136,  (2017).
\newblock \doi{10.1007/JHEP08(2017)136}.

\bibitem{Oblak:2016eij}
B.~Oblak.
\newblock \emph{{BMS Particles in Three Dimensions}}.
\newblock PhD thesis, Brussels U.,  (2016).

\bibitem{Stanford:2017thb}
D.~Stanford and E.~Witten, {Fermionic Localization of the Schwarzian Theory},
  \emph{JHEP}. {\bf 10}, \penalty0 008,  (2017).
\newblock \doi{10.1007/JHEP10(2017)008}.

\bibitem{Maldacena:1997re}
J.~M. Maldacena, {The Large N limit of superconformal field theories and
  supergravity}, \emph{Adv. Theor. Math. Phys.} {\bf 2}, \penalty0 231--252,
  (1998).
\newblock \doi{10.1023/A:1026654312961}.

\bibitem{Maldacena:1998u}
A.~Strominger, {AdS(2) quantum gravity and string theory}, \emph{JHEP}. {\bf
  01}, \penalty0 007,  (1999).
\newblock \doi{10.1088/1126-6708/1999/01/007}.

\bibitem{Spradlin:1999bn}
M.~Spradlin and A.~Strominger, {Vacuum states for AdS(2) black holes},
  \emph{JHEP}. {\bf 11}, \penalty0 021,  (1999).
\newblock \doi{10.1088/1126-6708/1999/11/021}.

\bibitem{Castro:2008ms}
A.~Castro, D.~Grumiller, F.~Larsen, and R.~McNees, {Holographic Description of
  AdS(2) Black Holes}, \emph{JHEP}. {\bf 11}, \penalty0 052,  (2008).
\newblock \doi{10.1088/1126-6708/2008/11/052}.

\bibitem{Castro:2014ima}
A.~Castro and W.~Song, {Comments on $\mathrm{AdS}_2$ Gravity} (11. 2014).

\bibitem{Maldacena:2016hyu}
J.~Maldacena and D.~Stanford, {Remarks on the Sachdev-Ye-Kitaev model},
  \emph{Phys. Rev. D}. {\bf 94}\penalty0 (10), \penalty0 106002,  (2016).
\newblock \doi{10.1103/PhysRevD.94.106002}.

\bibitem{Bagrets:2016cdf}
D.~Bagrets, A.~Altland, and A.~Kamenev,
  {Sachdev\textendash{}Ye\textendash{}Kitaev model as Liouville quantum
  mechanics}, \emph{Nucl. Phys. B}. {\bf 911}, \penalty0 191--205,  (2016).
\newblock \doi{10.1016/j.nuclphysb.2016.08.002}.

\bibitem{Hong:2019tsx}
J.~Hong, F.~Larsen, and J.~T. Liu, {The scales of black holes with nAdS$_{2}$
  geometry}, \emph{JHEP}. {\bf 10}, \penalty0 260,  (2019).
\newblock \doi{10.1007/JHEP10(2019)260}.

\bibitem{Kraus:2005vz}
P.~Kraus and F.~Larsen, {Microscopic black hole entropy in theories with higher
  derivatives}, \emph{JHEP}. {\bf 09}, \penalty0 034,  (2005).
\newblock \doi{10.1088/1126-6708/2005/09/034}.

\bibitem{Kraus:2005zm}
P.~Kraus and F.~Larsen, {Holographic gravitational anomalies}, \emph{JHEP}.
  {\bf 01}, \penalty0 022,  (2006).
\newblock \doi{10.1088/1126-6708/2006/01/022}.

\bibitem{Gupta:2008ki}
R.~K. Gupta and A.~Sen, {Ads(3)/CFT(2) to Ads(2)/CFT(1)}, \emph{JHEP}. {\bf
  04}, \penalty0 034,  (2009).
\newblock \doi{10.1088/1126-6708/2009/04/034}.

\bibitem{Balasubramanian:2009bg}
V.~Balasubramanian, J.~de~Boer, M.~M. Sheikh-Jabbari, and J.~Simon, {What is a
  chiral 2d CFT? And what does it have to do with extremal black holes?},
  \emph{JHEP}. {\bf 02}, \penalty0 017,  (2010).
\newblock \doi{10.1007/JHEP02(2010)017}.

\bibitem{Castro:2010vi}
A.~Castro, C.~Keeler, and F.~Larsen, {Three Dimensional Origin of AdS$_{2}$
  Quantum Gravity}, \emph{JHEP}. {\bf 07}, \penalty0 033,  (2010).
\newblock \doi{10.1007/JHEP07(2010)033}.

\bibitem{Bhattacharyya:2012wz}
S.~Bhattacharyya, B.~Panda, and A.~Sen, {Heat Kernel Expansion and Extremal
  Kerr-Newmann Black Hole Entropy in Einstein-Maxwell Theory}, \emph{JHEP}.
  {\bf 08}, \penalty0 084,  (2012).
\newblock \doi{10.1007/JHEP08(2012)084}.

\bibitem{Keeler:2014bra}
C.~Keeler, F.~Larsen, and P.~Lisbao, {Logarithmic Corrections to $N \geq 2$
  Black Hole Entropy}, \emph{Phys. Rev. D}. {\bf 90}\penalty0 (4), \penalty0
  043011,  (2014).
\newblock \doi{10.1103/PhysRevD.90.043011}.

\bibitem{Charles:2015eha}
A.~M. Charles and F.~Larsen, {Universal corrections to non-extremal black hole
  entropy in $ \mathcal{N}\ge 2 $ supergravity}, \emph{JHEP}. {\bf 06},
  \penalty0 200,  (2015).
\newblock \doi{10.1007/JHEP06(2015)200}.

\bibitem{Banerjee:2011jp}
S.~Banerjee, R.~K. Gupta, I.~Mandal, and A.~Sen, {Logarithmic Corrections to
  N=4 and N=8 Black Hole Entropy: A One Loop Test of Quantum Gravity},
  \emph{JHEP}. {\bf 11}, \penalty0 143,  (2011).
\newblock \doi{10.1007/JHEP11(2011)143}.

\bibitem{Sen:2014aja}
A.~Sen, {Microscopic and Macroscopic Entropy of Extremal Black Holes in String
  Theory}, \emph{Gen. Rel. Grav.} {\bf 46}, \penalty0 1711,  (2014).
\newblock \doi{10.1007/s10714-014-1711-5}.

\bibitem{Sen:2008vm}
A.~Sen, {Quantum Entropy Function from AdS(2)/CFT(1) Correspondence},
  \emph{Int. J. Mod. Phys. A}. {\bf 24}, \penalty0 4225--4244,  (2009).
\newblock \doi{10.1142/S0217751X09045893}.

\bibitem{Sen:2012dw}
A.~Sen, {Logarithmic Corrections to Schwarzschild and Other Non-extremal Black
  Hole Entropy in Different Dimensions}, \emph{JHEP}. {\bf 04}, \penalty0 156,
  (2013).
\newblock \doi{10.1007/JHEP04(2013)156}.

\bibitem{Heydeman:2020hhw}
M.~Heydeman, L.~V. Iliesiu, G.~J. Turiaci, and W.~Zhao, {The statistical
  mechanics of near-BPS black holes} (11. 2020).

\bibitem{Jevicki:2016bwu}
A.~Jevicki, K.~Suzuki, and J.~Yoon, {Bi-Local Holography in the SYK Model},
  \emph{JHEP}. {\bf 07}, \penalty0 007,  (2016).
\newblock \doi{10.1007/JHEP07(2016)007}.

\bibitem{Jevicki:2016ito}
A.~Jevicki and K.~Suzuki, {Bi-Local Holography in the SYK Model:
  Perturbations}, \emph{JHEP}. {\bf 11}, \penalty0 046,  (2016).
\newblock \doi{10.1007/JHEP11(2016)046}.

\bibitem{Turiaci:2017zwd}
G.~Turiaci and H.~Verlinde, {Towards a 2d QFT Analog of the SYK Model},
  \emph{JHEP}. {\bf 10}, \penalty0 167,  (2017).
\newblock \doi{10.1007/JHEP10(2017)167}.

\bibitem{Liu:2019niv}
J.~Liu and Y.~Zhou, {Note on global symmetry and SYK model}, \emph{JHEP}. {\bf
  05}, \penalty0 099,  (2019).
\newblock \doi{10.1007/JHEP05(2019)099}.

\bibitem{Maldacena:2019ufo}
J.~Maldacena and A.~Milekhin, {SYK wormhole formation in real time},
  \emph{JHEP}. {\bf 04}, \penalty0 258,  (2021).
\newblock \doi{10.1007/JHEP04(2021)258}.

\bibitem{Teitelboim:1983ux}
C.~Teitelboim, {Gravitation and Hamiltonian Structure in Two Space-Time
  Dimensions}, \emph{Phys. Lett. B}. {\bf 126}, \penalty0 41--45,  (1983).
\newblock \doi{10.1016/0370-2693(83)90012-6}.

\bibitem{Jackiw:1984je}
R.~Jackiw, {Lower Dimensional Gravity}, \emph{Nucl. Phys. B}. {\bf 252},
  \penalty0 343--356,  (1985).
\newblock \doi{10.1016/0550-3213(85)90448-1}.

\bibitem{Almheiri:2014cka}
A.~Almheiri and J.~Polchinski, {Models of AdS$_{2}$ backreaction and
  holography}, \emph{JHEP}. {\bf 11}, \penalty0 014,  (2015).
\newblock \doi{10.1007/JHEP11(2015)014}.

\bibitem{Jensen:2016pah}
K.~Jensen, {Chaos in AdS$_2$ Holography}, \emph{Phys. Rev. Lett.} {\bf
  117}\penalty0 (11), \penalty0 111601,  (2016).
\newblock \doi{10.1103/PhysRevLett.117.111601}.

\bibitem{Forste:2017kwy}
S.~Forste and I.~Golla, {Nearly AdS$_2$ sugra and the super-Schwarzian},
  \emph{Phys. Lett. B}. {\bf 771}, \penalty0 157--161,  (2017).
\newblock \doi{10.1016/j.physletb.2017.05.039}.

\bibitem{Cvetic:2016eiv}
M.~Cveti\v{c} and I.~Papadimitriou, {AdS$_{2}$ holographic dictionary},
  \emph{JHEP}. {\bf 12}, \penalty0 008,  (2016).
\newblock \doi{10.1007/JHEP12(2016)008}.
\newblock [Erratum: JHEP 01, 120 (2017)].

\bibitem{Nayak:2018qej}
P.~Nayak, A.~Shukla, R.~M. Soni, S.~P. Trivedi, and V.~Vishal, {On the Dynamics
  of Near-Extremal Black Holes}, \emph{JHEP}. {\bf 09}, \penalty0 048,  (2018).
\newblock \doi{10.1007/JHEP09(2018)048}.

\bibitem{Moitra:2018jqs}
U.~Moitra, S.~P. Trivedi, and V.~Vishal, {Extremal and near-extremal black
  holes and near-CFT$_{1}$}, \emph{JHEP}. {\bf 07}, \penalty0 055,  (2019).
\newblock \doi{10.1007/JHEP07(2019)055}.

\bibitem{Saad:2019lba}
P.~Saad, S.~H. Shenker, and D.~Stanford, {JT gravity as a matrix integral} (3.
  2019).

\bibitem{Iliesiu:2019xuh}
L.~V. Iliesiu, S.~S. Pufu, H.~Verlinde, and Y.~Wang, {An exact quantization of
  Jackiw-Teitelboim gravity}, \emph{JHEP}. {\bf 11}, \penalty0 091,  (2019).
\newblock \doi{10.1007/JHEP11(2019)091}.

\bibitem{Mertens:2019tcm}
T.~G. Mertens and G.~J. Turiaci, {Defects in Jackiw-Teitelboim Quantum
  Gravity}, \emph{JHEP}. {\bf 08}, \penalty0 127,  (2019).
\newblock \doi{10.1007/JHEP08(2019)127}.

\bibitem{Stanford:2019vob}
D.~Stanford and E.~Witten, {JT gravity and the ensembles of random matrix
  theory}, \emph{Adv. Theor. Math. Phys.} {\bf 24}\penalty0 (6), \penalty0
  1475--1680,  (2020).
\newblock \doi{10.4310/ATMP.2020.v24.n6.a4}.

\bibitem{Iliesiu:2019lfc}
L.~V. Iliesiu, {On 2D gauge theories in Jackiw-Teitelboim gravity} (9. 2019).

\bibitem{Iliesiu:2020qvm}
L.~V. Iliesiu and G.~J. Turiaci, {The statistical mechanics of near-extremal
  black holes}, \emph{JHEP}. {\bf 05}, \penalty0 145,  (2021).
\newblock \doi{10.1007/JHEP05(2021)145}.

\bibitem{Almheiri:2020cfm}
A.~Almheiri, T.~Hartman, J.~Maldacena, E.~Shaghoulian, and A.~Tajdini, {The
  entropy of Hawking radiation}, \emph{Rev. Mod. Phys.} {\bf 93}\penalty0 (3),
  \penalty0 035002,  (2021).
\newblock \doi{10.1103/RevModPhys.93.035002}.

\bibitem{Raju:2020smc}
S.~Raju, {Lessons from the Information Paradox} (12. 2020).

\bibitem{Cotler:2016fpe}
J.~S. Cotler, G.~Gur-Ari, M.~Hanada, J.~Polchinski, P.~Saad, S.~H. Shenker,
  D.~Stanford, A.~Streicher, and M.~Tezuka, {Black Holes and Random Matrices},
  \emph{JHEP}. {\bf 05}, \penalty0 118,  (2017).
\newblock \doi{10.1007/JHEP05(2017)118}.
\newblock [Erratum: JHEP 09, 002 (2018)].

\bibitem{Saad:2018bqo}
P.~Saad, S.~H. Shenker, and D.~Stanford, {A semiclassical ramp in SYK and in
  gravity} (6. 2018).

\end{thebibliography}

\end{document}